\documentclass[twocolumn,aps,prb,showkeys,superscriptaddress,letterpaper]{revtex4-2}
\usepackage[utf8]{inputenc}
\usepackage{amsmath}
\usepackage{amssymb}
\usepackage{graphicx}
\graphicspath{{plots/}}
\usepackage[usenames,dvipsnames]{xcolor}

\usepackage[unicode=true,
 bookmarks=true,bookmarksnumbered=false,bookmarksopen=false,
 breaklinks=false,pdfborder={0 0 0},pdfborderstyle={},backref=false,colorlinks=true,]
 {hyperref}
\hypersetup{linkcolor=blue,citecolor=blue,urlcolor=blue}

\begin{document}


\title{Topological phases of strongly-interacting time-reversal invariant topological superconducting chains under a magnetic field}

\author{Leandro M. Chinellato}
\author{Claudio J. Gazza}

\affiliation{Instituto de F\'\i sica Rosario, CONICET, and Facultad de Ciencias Exactas, Ingenier\'\i a y Agrimensura,
\\
Universidad Nacional de Rosario, 2000 Rosario Argentina}

\author{Alejandro M. Lobos}\email{alejandro.martin.lobos@gmail.com}

\affiliation{Instituto Interdisciplinario de Ciencias B\'{a}sicas - Consejo Nacional de Investigaciones Cient\'{i}ficas y T\'{e}cnicas }
\affiliation{Facultad de Ciencias Exactas y Naturales - Universidad Nacional de Cuyo}

\author{Armando A. Aligia}
\affiliation{Instituto de Nanociencia y Nanotecnolog\'{\i}a CNEA-CONICET, GAIDI,
Centro At\'{o}mico Bariloche and Instituto Balseiro, 8400 Bariloche, Argentina}

\begin{abstract}
Using the density-matrix renormalization group, we 
determine the different topological phases and low-energy 
excitations of a time-reversal invariant topological superconducting (TRITOPS) wire with extended $s$-wave superconductivity, Rashba spin-orbit coupling (SOC) and on-site repulsion $U$, under an externally applied Zeeman field $J$. For the case in which $J$ is perpendicular to the SOC,
the model describes a chain of Shiba impurities on top of a superconductor with extended superconductor pairing. 
We identify the different topological phases of the model at temperature $T=0$, and in particular study the stability of the TRITOPS phase against the Zeeman field $J$ and the chemical potential $\mu$, for different values of $U$. 
In the case where the magnetic field $J$ is perpendicular to the SOC axis,
the pair of Kramers-degenerate Majorana zero modes  at the edges of the system that exist for $J=0$, remain degenerate
until a critical value of the magnetic field is reached. 
For $J$ parallel to the SOC and up to moderate values of $U$, the fractional spin projection 
$\langle S_y \rangle=1/4$ at the ends, found for non-interacting wires at $U=0$, is recovered. In addition, the analytic expression
that relates $\langle S_y \rangle$ with $J$ for finite non-interacting
chains is shown to be universal up to moderate values of $U$. 
\end{abstract}

\maketitle

\section{Introduction}
The quest for topological phases of matter and, in particular, topological superconductors (TOPS) has been a major pursuit in condensed matter physics for the last 20 years \cite{kitaev2001}. TOPS phases hosting elusive Majorana zero-modes (MZMs) have attracted a lot of interest both from the fundamental point of view, as well as for potential uses in fault-tolerant quantum computation due to their exotic non-Abelian anyonic statistics   \cite{Nayak08_RMP_Topological_quantum_computation}. 

Although much of the progress in this area has been achieved within a  framework of non-interacting electrons (i.e.,  the topological classification of TOPS phases according to their symmetries and the identification of possible topological invariants), the effects of interactions still remains as a conceptually important open question. 
Moreover, many of the technologically relevant applications might involve low-dimensional TOPS systems, for which the effects of interactions are enhanced \cite{giamarchi}. Therefore, the study of interaction effects on TOPS is also relevant from the technological perspective.

Up to now, a variety of different physical systems have been proposed to realize TOPS phases hosting MZM states:  $\nu= 5/2$ fractional
quantum Hall state \cite{Moore91_FQH_state_5-2},  superfluid He-3 \cite{Salomaa88_Cosmiclike_DW_in_He3}, proximitized topological insulator-superconductor structures \cite{Fu08_Proximity-effect_and_MF_at_the_surface_of_TIs}, superconducting heterostructures combining proximity-induced superconductivity, semiconductors with strong Rashba spin-orbit interaction and Zeeman fields \cite{Sau10_Proposal_for_MF_in_semiconductor_heterojunction, Lutchyn10_MF_and_Topological_transition_in_SM_SC_Heterostructures, Oreg10_Helical_liquids_and_MF_in_QW}, etc. All these systems are potential realizations of TOPS phases which break time-reversal symmetry 
(\textquotedblleft class D \textquotedblright TOPS in the Altland-Zirnbauer classification \cite{Altland97_Symmetry_classes, Ryu10_Topological_classification}). 

A different class, the time-reversal invariant TOPS (TRITOPS) or DIII class TOPS originally proposed by Qi \textit{et al} \cite{Qi10_TRITOPS, Qi10_Topological_invariants_for_TRITOPS}, has been predicted by Zhang, Kane and Mele (ZKM) to arise in 1D or 2D geometries by combining 
semiconductors with strong Rashba spin-orbit coupling (SOC)
(i.e., nanowires or films) proximitized with 
extended $s$-wave superconductors \cite{Zhang13_TRITOPS}. 
The TRITOPS have been recently the subject of intense theoretical research \cite{Qi10_TRITOPS, Qi10_Topological_invariants_for_TRITOPS,Zhang13_TRITOPS,
deng12,dumi13,kese13,haim14,mellars16,camjayi17,ali18,schrade18,arra19,casas19,ali19,haim19,gabriel22}.
For 1D TRITOPS, a key feature is the existence
of Kramers pairs of MZMs at the edges of the system.
Another peculiar feature is that the spin projection 
at the ends in the direction of the SOC is $\pm 1/4$ 
\cite{Qi10_TRITOPS,ali18,ali19}. For a TRITOPS wire of length $L$, MZMs are well defined as long as $L\gg \xi$, with $\xi$ the MZM localization  length. Under these conditions, an external magnetic field applied to \textit{one half} of the wire in the direction of the SOC produces a Zeeman-split pair of low-energy MZMs with total spin projection at the end equal to 1/4 or -1/4, depending on the sign
of the magnetic field \cite{ali18,ali19}.

The effect of repulsive interactions in 1D TOPS and TRITOPS
has been studied  in previous works using e.g., mean-field approaches \cite{Danon15_Interaction_effects_on_SM_NWs}, 
density-matrix renormalization group (DMRG) \cite{Haim14_DMRG_study_of_interacting_TRITOPS}, the Abelian bosonization framework \cite{Haim16_Interaction_driven_TSC_in_1D}, and 
numerical renormalization group for two sites \cite{ma2023}. While for 1D TOPS interactions tend to weaken the superconducting correlations, therefore weakening the TOPS phase \cite{Gangadharaiah11_Majorana_fermions_in_1D_interacting_wires, Lobos12_Interplay_disorder_interaction_Majorana_wire}, it was suggested that repulsive interactions in a 1D system stabilizes the TRITOPS phase. The basic stabilization mechanism consists in local repulsive interactions which penalize the proximity-induced singlet pairing  with respect to the proximity-induced triplet pairing \cite{Sun14_Tuning_bewteen_singlet_triplet_pairing_with_interactions, Haim16_Interaction_driven_TSC_in_1D}. In addition, although it is not the scope of the present work, we mention in passing that the effect of attractive interactions on TRITOPS
has also been studied \cite{Keselman15_TRITOPS_phase_in_attractive_1D_fermion_systems}.

In this article
we explore the effects of on-site repulsive interaction $U$ on 
the ZKM model in the presence of a magnetic field $J$.
For $J$ perpendicular to the SOC, the model describes
hybrid magnet-superconductor systems with TOPS and TRITOPS phases, in particular  
magnetic adatoms (i.e., Fe, Co, or Mn atoms) deposited on top of  a superconductor, a system usually known as a 
\textquotedblleft Shiba chain \textquotedblright. 
Recent experimental progress in these type of hybrid nanostructures have shown preliminary evidence of MZMs in the $dI/dV$ STM signal \cite{Nadj-Perdge13_Majorana_fermions_in_Shiba_chains, Klinovaja13_TSC_and_Majorana_Fermions_in_RKKY_Systems, Braunecker13_Shiba_chain, NadjPerge14_Observation_of_Majorana_fermions_in_Fe_chains, Ruby15_MBS_in_Shiba_chain, Pawlak16_Probing_Majorana_wavefunctions_in_Fe_chains,Kim18_MBS_in_Shiba_chains}, drawing a lot of interest. However, the small size of the parent superconductor gap (typically Pb) imposes practical difficulties in all type of  proximity-induced TOPS, such as e.g., stringent low-temperature requirements and limited spectral resolution of the experiments. For this reason, recent theoretical proposals have put forward the possibility to observe both TOPS and TRITOPS in nanostructures made of magnetic impurities deposited at the surface of unconventional high $T_c$ superconductors, generating renewed interest on these hybrid structures \cite{Crawford20_MZM_in_high_Tc_hybrids}.

In this work, using the density-matrix renormalization group (DMRG) method, we study the topological phase diagram of the system for 
finite magnetic field perpendicular to the SOC and for different values of $U$. 
We also explore the response of the MZMs to the presence of a magnetic field applied to one half of the chain. Such a magnetic probe can help to detect and identify the topological phase of the chain. In particular, we show that the fractional spin 1/4 excitations at each end of the wire, predicted to emerge in  non-interacting models
for magnetic field parallel to the SOC \cite{Qi10_TRITOPS,ali18,ali19}, are robust to the presence of strong interactions.

The paper is organized as follows. In Section \ref{model}
we explain our model. Section \ref{res} contains the main results and Section \ref{sum} is a summary and discussion.

\section{Theoretical Model and Methods}

\label{model}

We consider the following discrete Hamiltonian encoding the minimal  
ingredients leading to a TRITOPS phase (in close analogy to the ZKM model in the continuum), with additional Zeeman and an on-site repulsion terms:

\begin{eqnarray}
H &=& \sum_{j} \left[\left( t  \mathbf{c}_j^\dagger \mathbf{c}_{j+1} - \left(\frac{\mu}{2}+\frac{U}{4}\right)\mathbf{c}_j^\dagger \mathbf{c}_j +  \Delta  c_{j,\uparrow}^\dagger c_{j+1,\downarrow}^\dagger \right. \right.  
\nonumber \\
&&\left. \left. + i\alpha_R  \mathbf{c}_j^\dagger \sigma_y \mathbf{c}_{j+1}+ \text{H.c.}\right) - J\mathbf{c}_j^\dagger \sigma_\beta \mathbf{c}_{j} + U n_{j,\uparrow}n_{j,\downarrow}\right ]
\label{ham}
\end{eqnarray}
where $\mathbf{c}^\dagger_j=(c^\dagger_{j,\uparrow},c^\dagger_{j,\downarrow})$ is a spinor containing both fermionic creation operators at site $j$ with spin projections $\{\uparrow, \downarrow \}$,  $\sigma_\alpha$ (with $\alpha=\{x,y,z\}$) are the $2\times 2$ Pauli matrices,
$t$ is the nearest-neighbor hopping amplitude, $\mu$ is the chemical potential, $\alpha_R$ is the Rashba SOC in the 
$y$ direction, $\Delta$ is the extended $s$-wave amplitude of the superconducting first-neighbor pairing correlations, 
and $U$ is the on-site electron-electron repulsion. This particular form of Eq. (\ref{ham}) ensures that  for any value of $U$, $\mu=0$  corresponds to the particle-hole symmetric point.

While the presence of a nearby bulk superconductor usually screens the electron-electron interaction, in low-dimensional nanostructures of reduced dimensions, local repulsion terms of this type might be relevant, and in fact (as we show below) this is the case for the ground-state phase diagram of this system. In the above model, the Zeeman parameter $J$ can either represent  the effect of an external magnetic field  ($J=\mu_B B)$ applied along the  
$\beta$ direction, or (in the case of atomic Shiba chains) the effect of a local exchange field originated in a microscopic \textit{s-d} exchange interaction $I_{sd}$
($J=2I_{sd} S^j_z$) between the 
conduction states and the  magnetic impurities $S^j_z$ assumed ferromagnetically aligned along $z$ at each 
site $j$ of the chain. In this case, $\beta=z$ and the effect is 
similar to a magnetic field perpendicular to the SOC.
We will also consider the case when $J$ is parallel to the 
SOC ($\beta=y$).

All the numerical results presented in this work have been obtained by the means of DMRG computations, implemented using the ITensor software library \cite{10.21468/SciPostPhysCodeb.4}. We have implemented the necessary maximum bond dimension (400 in the worst case) which allowed us to keep the truncation error cutoff of $10^{-10}$ throughout.

\section{Results}
\label{res}

\subsection{Topological phase diagram}
\label{tpd}

\begin{figure}[htbp]
    \centering \includegraphics[width=0.52\columnwidth, viewport=170 0 350 850]{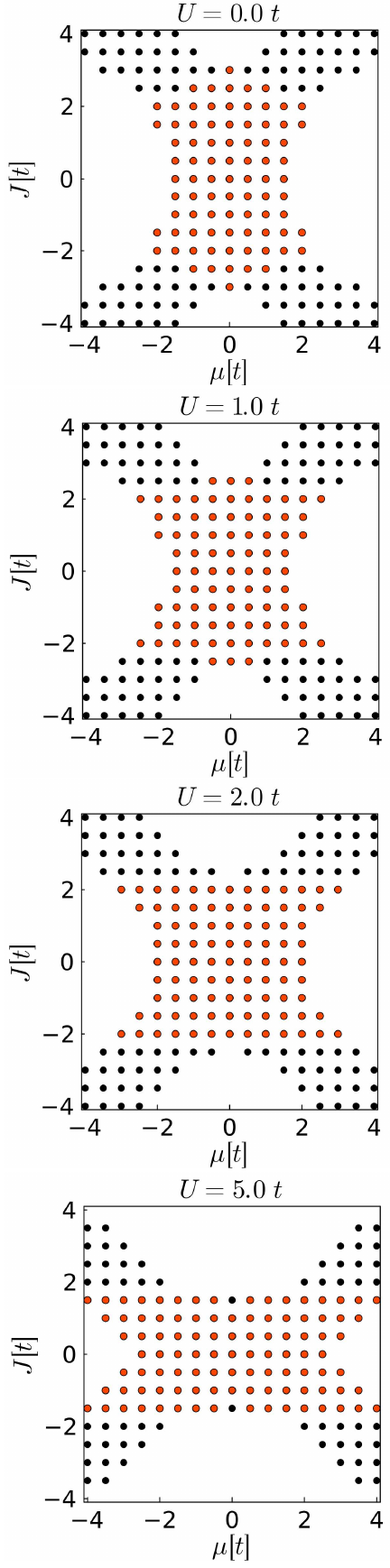}
\caption{(Color online) Topological phase diagram as a function of chemical potential and magnetic field perpendicular to the SOC. Red (black) dots correspond to four-fold (two-fold) degeneracy of the entanglement spectrum. Parameters are $\Delta=1.2$ 
and $\alpha=0.8$, and the length of the chain is $L=400$ sites.}  
\label{diag}
\end{figure}

We now focus on the ground-state properties of the system and study the topological phase diagram as a function of  the chemical potential $\mu$ and the Zeeman field $J$ perpendicular to the SOC, for different values of the interaction parameter $U$ (see Fig. \ref{diag}). The other parameters of the model are fixed throughout the rest of the paper to the values $\Delta=1.2$, $\alpha_R=0.8$ (here the hopping amplitude $t=1$ has been chosen as the unit of energy). This particular parameter set  has been chosen to coincide with those used in Ref. \cite{Crawford20_MZM_in_high_Tc_hybrids}.

We determine the topological nature of the ground state by analyzing the degeneracy of the reduced density matrix entanglement spectrum \cite{Pollmann10_Entaglement_spectrum_of_topological_phases_in_1D, Turner11_Topological_phases_of_1D_fermions}. Given a quantum system which can be divided into two subsystems $A$ and $B$, the entanglement spectrum is the spectrum of eigenvalues of the reduced density-matrix $\rho_A\ (\rho_B)$, obtained after tracing out 	the $B$ ($A$) degrees of freedom. The change of degeneracies in the entanglement spectrum is indicative of topological quantum phase transitions occurring in the ground state of the whole system, and is 
related to the degeneracy of the ground-state and the number of MZMs per end of the chain \cite{Pollmann10_Entaglement_spectrum_of_topological_phases_in_1D, Turner11_Topological_phases_of_1D_fermions}. 

Generically speaking, starting from a parameter regime which realizes a time-reversal symmetric superconductor [$J=0$ in our model Eq. (\ref{ham})] and for low or moderate values of $\mu$, we obtain a 4-fold degenerate ground state indicative of a TRITOPS phase (red dots in Fig. \ref{diag}). Interestingly, we see that this phase is robust against the effect of a uniform magnetic field perpendicular to the SOC, and only for quite large values of $J$ beyond a critical line $J_c(\mu)$ the system becomes a DIII TOPS with a two-fold degenerate ground state (see black dots). Additionally, for extremely low (large) values of $\mu$, the bands can be completely depleted (filled) and the system becomes a trivial insulator with a non-degenerate ground state (white region in Fig. \ref{diag}). 

The aforementioned robustness of the 4-fold ground-state multiplet is quite surprising given the fact that time-reversal symmetry no longer protects the TRITOPS phase. In the non-interacting case, this is related to the presence of an additional chiral symmetry, implemented by the operator $\mathcal{S}=\sigma_y\tau_y$ (where the Pauli matrices $\tau_\alpha$ operate on the Nambu space) which anticommutes with $H$. Indeed, for $U=0$, and taking periodic boundary conditions in Eq. (\ref{ham}) , the Hamiltonian matrix of the system takes the compact form in $k$-space 
$\mathcal{H}_k=(\epsilon_k -\mu)\sigma_0 \tau_z+\alpha \sigma_y \tau_z + \Delta_k \sigma_0 \tau_x +J \sigma_\beta \tau_0$, where the Nambu basis $\Psi_k=(c_{k\uparrow},c_{c,\downarrow},c^\dagger_{-k\downarrow}, -c^\dagger_{-k\uparrow})^T$ has been used, and where $\epsilon_k=2t\cos(k)$ and $\Delta_k=\Delta \cos(k)$. It is easy to see that when $\beta=z$ the chiral operator $\mathcal{S}$ anticommutes with $\mathcal{H}_k$, and generates a chiral symmetry which is \textit{additional} to the time-reversal symmetry occurring for $J=0$. This additional symmetry allows to compute a $Z$ invariant which counts the number of MZMs at each end of the wire \cite{ Dumitrescu14_Additional_chiral_symmetry_on_TRITOPS_in_1D}.

\begin{figure} 
 \includegraphics[width=\columnwidth]{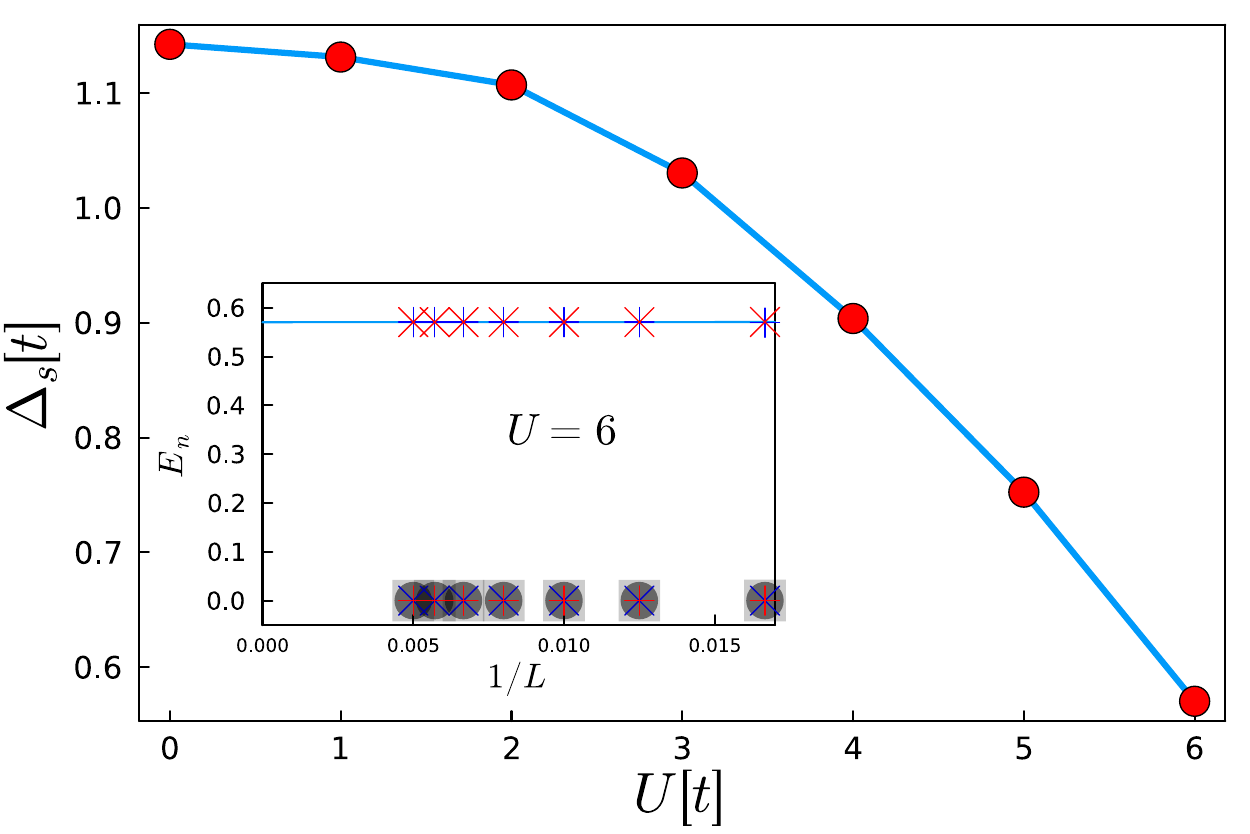}
\caption{(Color online) Superconducting single-particle excitation gap $\Delta_s$ (defined in the main text) computed for several values of $U$, and 
for $\mu=J=0$. Inset: Finite-size scaling showing the first excitations in the odd-parity subspace for $U=6$. }
\label{gaps}-
\end{figure}

On the other hand, as can be seen in Fig. \ref{diag}, 
the presence of a local on-site interaction Hubbard $U$ term has an important effect on the topological phase diagram, as it tends to weaken the TRITOPS phase with respect to the magnetic field. This effect can be qualitatively understood in terms of an effective non-interacting model with a smaller renormalized superconducting bulk gap due to the repulsive interaction. In the following we denote this gap excluding the low-energy
excitations related with the MZMs as $\Delta_s$.
This gap is calculated as follows. The one-particle excitations energies are defined as the different energies in the subspace with odd number of particles minus the ground-state energy (which lies in the subspace with even number of particles) 
\begin{equation}
E_n=E^\text{(odd)}_{n}-E^\text{(even)}_{g},
\label{en}
\end{equation}
where the subscript $g$ denotes the ground state.
Among these $E_n$, in the topologically non-trivial regions,  there is a  low-lying subset which correspond to the MZMs, with a small exponential splitting $\sim e^{-L/\xi}$ for a finite chain, due to the mixing of the MZMs between both ends. The corresponding excitation energies of this subset behave as $E_n \rightarrow 0$ for $L \rightarrow \infty$, and can be easily identified with a finite-size scaling analysis. The next excitation energy above this multiplet defines $\Delta_s$, which can be identified with bulk excitations. In Fig. \ref{gaps}, we show  
$\Delta_s$ as a function of on-site repulsion. One can clearly see that the value of $\Delta_s$ decreases by nearly a factor 2 as $U$ increases from 2 to 6. 
The detrimental effects of the repulsive interactions on 
$\Delta_s$ allows to qualitatively understand the topological phase diagram on Fig. \ref{diag}. In the inset of Fig. \ref{gaps}, we show the aforementioned low-lying mulltiplet of MZMs, and the bulk-excitation gap $\Delta_s$ (for which no appreciable dependence of $\Delta_s \sim 0.58$ on the length of the chain $L$ is observed), computed for the particular value $U=6$.

This renormalization of $\Delta_s$ due to the repulsion $U$ has detrimental effects on the stability of the TRITOPS phase, in particular when $J$ is increased (see Fig. \ref{diag}). Note however that  increasing $U$ also strengthens the TRITOPS phase with respect to the chemical potential $\mu$. This effect might be actually beneficial for potential implementations of TRITOPS in devices, as it expands the parameter regime near the line $J=0$ where this phase is realized. Indeed, by changing the parameters of the model, we can go from either a TRITOPS  with 4-fold degeneracy of the ground state, to a D-class topological phase with 2-fold degeneracy, to a trivial superconducting phase with a non-degenerate ground state. 

As a way to characterize the different phases of the model, in the next sections we consider an inhomogeneous Zeeman term applied to one half of the system (i.e., the left half). The introduction of a time-reversal symmetry-breaking interaction to only one end of the system allows to phenomenologically characterize the behaviour of the MZMs arising in TRITOPS.



\subsection{Magnetic field at the end perpendicular to the SOC}
\label{per}

\begin{figure}[h]
    \centering
    \includegraphics[width=0.8\columnwidth]{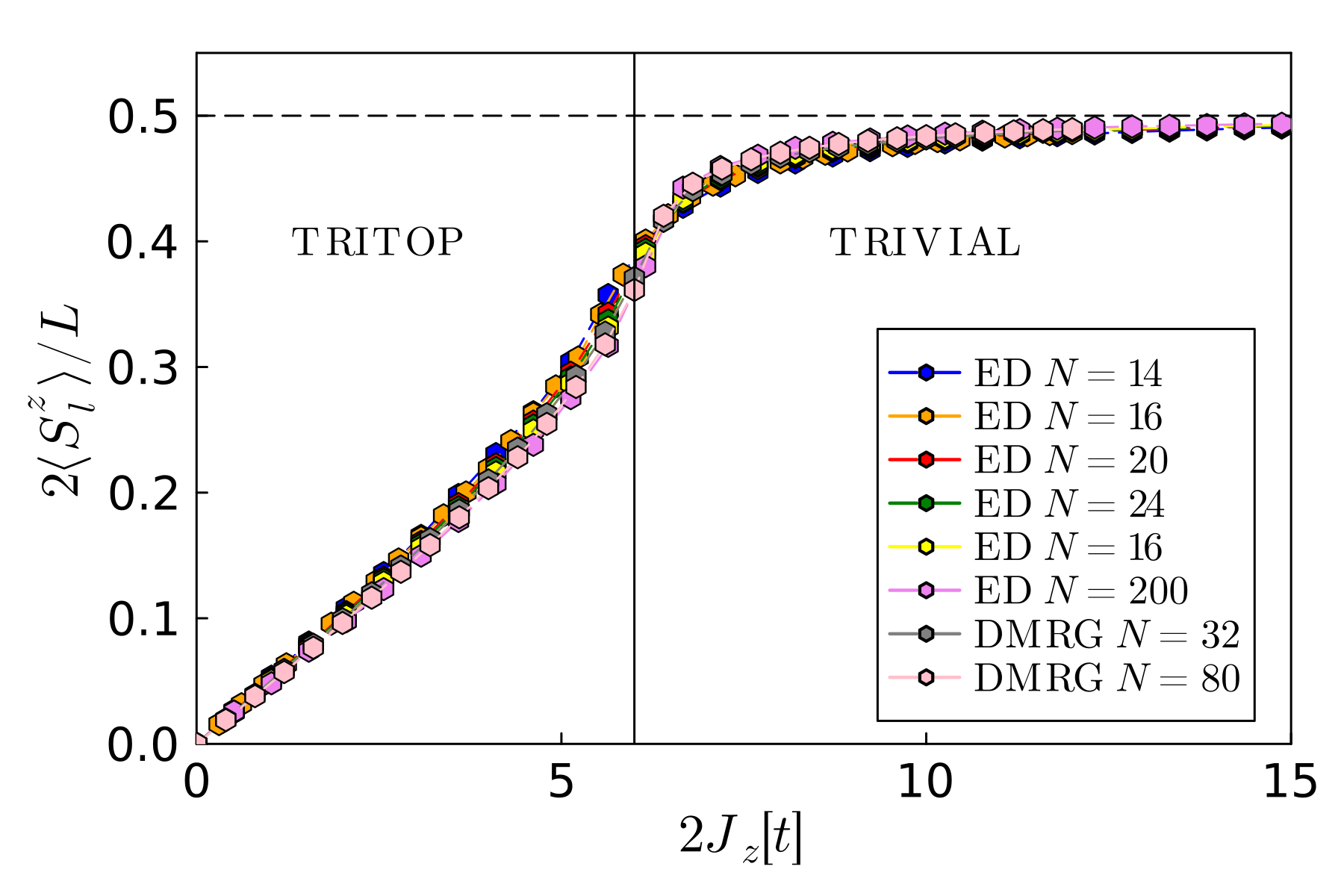}
    \caption{(Color online) Magnetization at the left side of the non-interacting ($U=0$) chain as a function of magnetic field applied to the left half of the chain. The different data sets represent different lengths $L$. The parameters set is  $t =1$, $\Delta=1.2$, $\alpha=0.8$ and $\mu=U=0$.}
    \label{per0}
\end{figure}

In this section we explore the fate of the MZMs in the TRITOPS phase when 
a magnetic field perpendicular to the SOC is applied to
the left part of the chain [Eq. (\ref{ham}) with 
a term 
$-J\sum_{j=1}^{L/2} \textbf{c}_j^{\dag}\sigma_{z}\textbf{c}_j$].
To that end we begin with the study of the non-interacting case $U=0$, where exact calculations independent of the DMRG procedure are available, and compute the magnetization of the left 
half of the chain $S_l^{z}=\sum_{j=1}^{L/2}S_i^z$.

In the topological phases and for low values of $J$, one expects the magnetization 
to be dominated by the MZMs, which are localized near both ends of the chain. For this reason, the spatial extension of the magnetic field is not important as long as it is longer than the MZM localization length $\xi$. However, rather surprisingly, for a magnetic field perpendicular to the SOC, 
the Kramers-degenerate MZMs of the TRITOPS are not
mixed by the magnetic field \cite{ali18}. This is related to the additional chiral symmetry $\mathcal{S}=\sigma_y\tau_y$ mentioned above. The leading correction to the ground-state energy becomes  of 
second-order in $J$ 
(i.e., quasiparticles are excited into the bulk and then return to the ground state),  leading to a linear 
dependence of $\langle S_l^{z} \rangle$ with $J$. 
This is in fact the behavior observed for small $J$,  displayed in Fig. \ref{per0}. The slope is a fraction of 
$J/\Delta_s$, where $\Delta_s$ is the superconducting gap, as 
expected for bulk excitations. For $U=0$, the model can be solved exactly without
using DMRG and we used these calculations to check our DMRG results. Note that all the plots eventually saturate at the value 1/2, corresponding to the completely polarized ground state, as is physically expected for very large values of the magnetic field.

\begin{figure}[h]
    \centering
    \includegraphics[width=0.8\columnwidth]{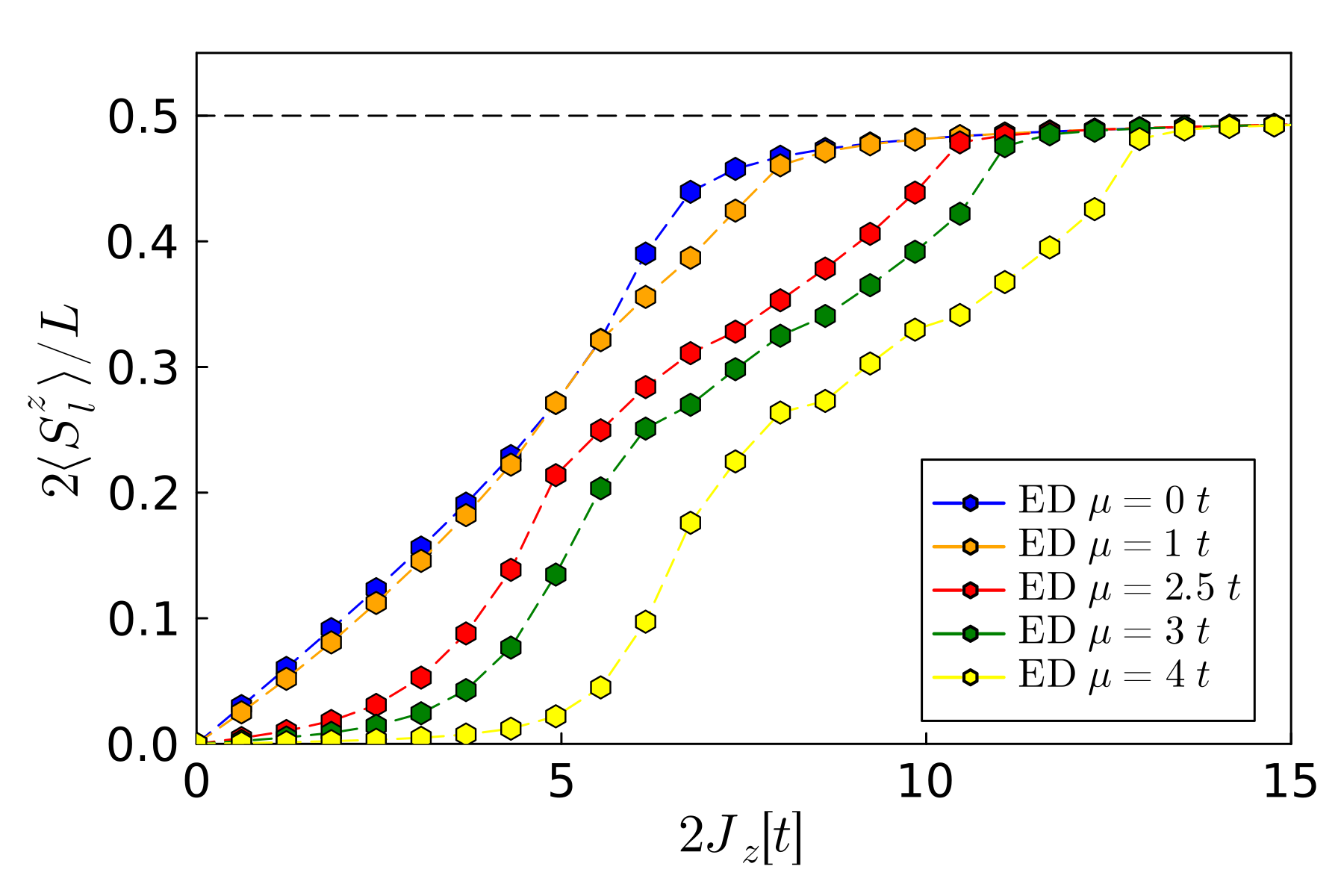}
    \caption{(Color online) Magnetization of the left half of the chain for $L=32$ sites and several values of 
    $\mu$. Other parameters as in Fig. \ref{per0}.}
    \label{permu}
\end{figure}

In Fig. \ref{permu} we show the changes introduced by a variation of the chemical potential. For 
$\mu \geq 2$ the system enters the trivial phase and the MZMs disappear. Therefore, the effect of the magnetic field is much weaker for small $J$. However, for large $\mu$, increasing $J$
the system enters the topological phase with one MZM at each end (black dots in Fig. \ref{diag}) and 
$\langle S_l^{z} \rangle$ increases in that region (for example for $2.6 < J/t < 9.4$ for $\mu=2.5$) before re-entering 
the trivial phase for large $J$, where $\langle S_l^{z} \rangle$ saturates at the value 1/2.

\begin{figure}[h]
    \centering
    \includegraphics[width=0.8\columnwidth]{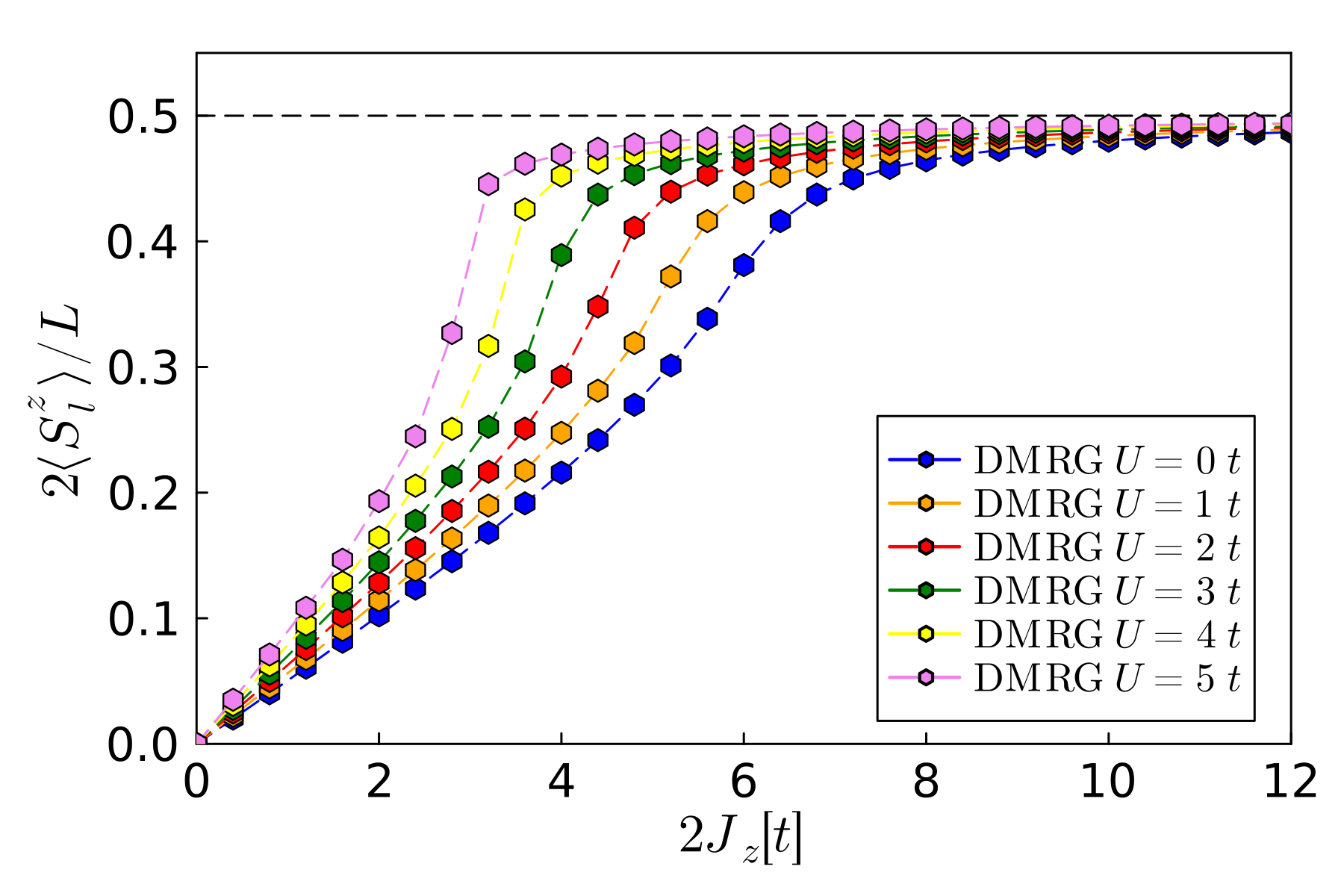}
    \caption{(Color online) Magnetization of the left half of the chain for $L=20$ sites and several values of 
    $U$. Other parameters as in Fig. \ref{per0}.}
    \label{peru}
\end{figure}

Finally, in Fig. \ref{peru} we explore the effect of a finite $U$.
As discussed in Section \ref{diag}, the gap $\Delta_s$ 
decreases with increasing $U$ and therefore the slope
of $\langle S_l^{z} \rangle$ for small $J$, which is expected to be inversely proportional to $\Delta_s$ increases. This behaviour confirms our interpretation of a strongly interacting TRITOPS chain in terms of an effectively non-interacting TRITOPS with a renormalized parameter $\Delta_s$.

\subsection{Magnetic field at the end parallel to the SOC}
\label{para}

For a very long chain in the regime $L\gg \xi$, and magnetic field parallel to 
the SOC, the system effectively behaves as if free fractionalized spins 1/4 
existed at each end. This picture is based on the fact that  an infinitesimally small $J$
generates a magnetisation $\langle S_l^{y}\rangle =1/4$, where 
$S_l^{y}=\sum_{i=1}^{L/2}S_i^y$ under a magnetic field 
$-J\sum_{j=1}^{L/2} \textbf{c}_j^{\dag}\sigma_{y}\textbf{c}_j$
applied to the left half of the chain \cite{Qi10_TRITOPS,ali18,ali19}.
On the other hand, for a finite non-interacting chain,  due to the mixing of MZMs at the ends, the lowest-lying Kramers-degenerate one-particle excitations have a small but finite energy $E(0)$ for $J=0$, which decays exponentially
with the length of the chain.
For a finite Zeeman energy $J$, the Kramers degeneracy is 
broken and $E(J)$, which corresponds to $E_1$ in Eq. (\ref{en}), decreases.
$E(J)$ has been calculated
analytically in Ref. \onlinecite{ali18}, and the expectation value of the 
spin projection is described by the simple expression
\cite{ali18}

\begin{equation}
    \langle S_{l}^{y} \rangle = \frac{2J_y}{4\sqrt{(2J_y)^2 + 16E(0)^2}}.
    \label{sy}
\end{equation}
Therefore, the magnetization increases and saturates to the value $1/4$ with an applied field
which is orders of magnitude smaller than in the case of a magnetic field perpendicular to the SOC
discussed in Section \ref{per}.

\begin{figure}[h]
    \centering
    \includegraphics[width=0.8\columnwidth]{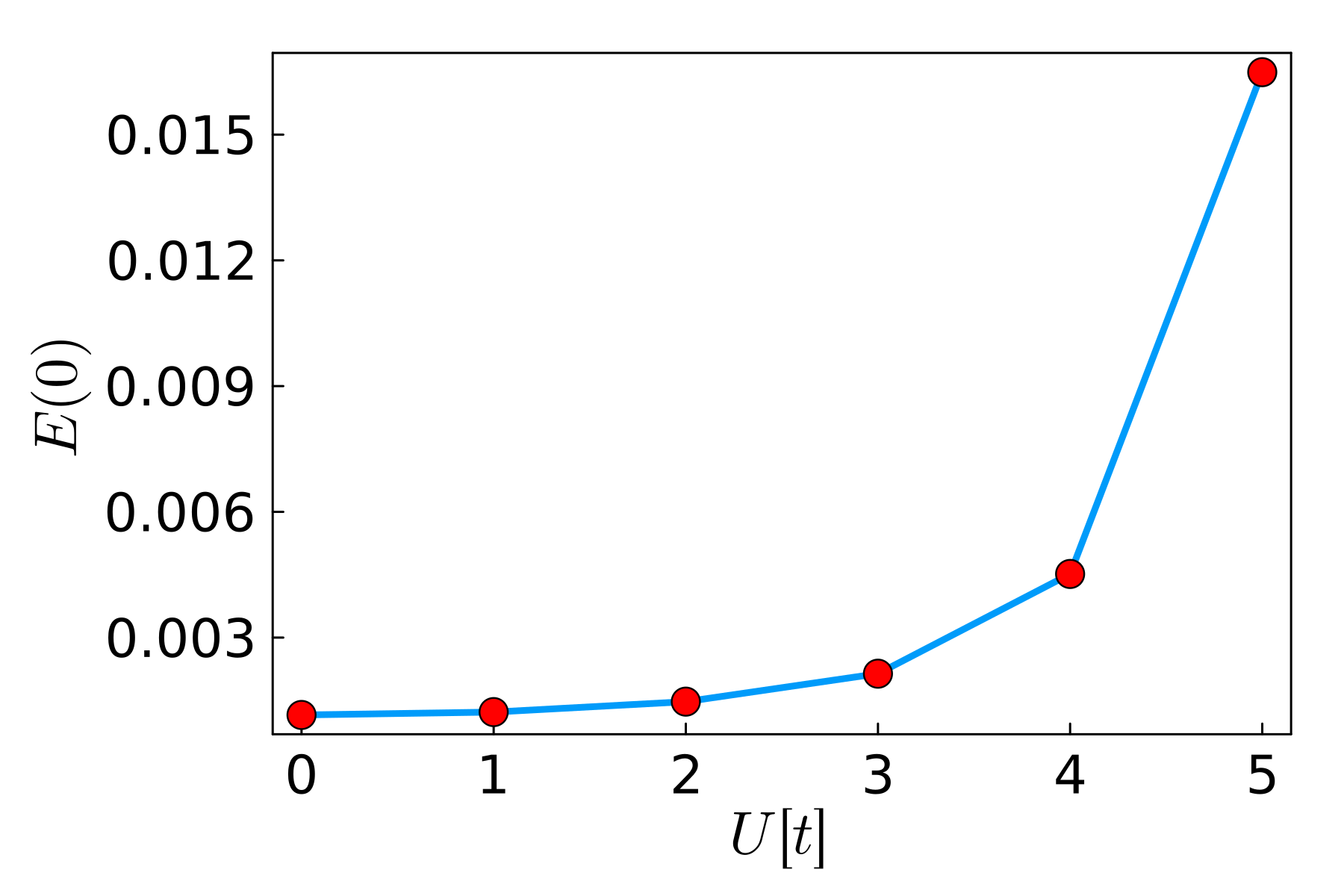}
    \caption{(Color online) Energy of the lowest one-particle excitation as function of $U$ for a chain of $L=20$ sites.
    Other parameters as in Fig. \ref{per0}.}
    \label{e0u}
\end{figure}

\begin{figure}[h]
    \centering
    \includegraphics[width=\columnwidth]{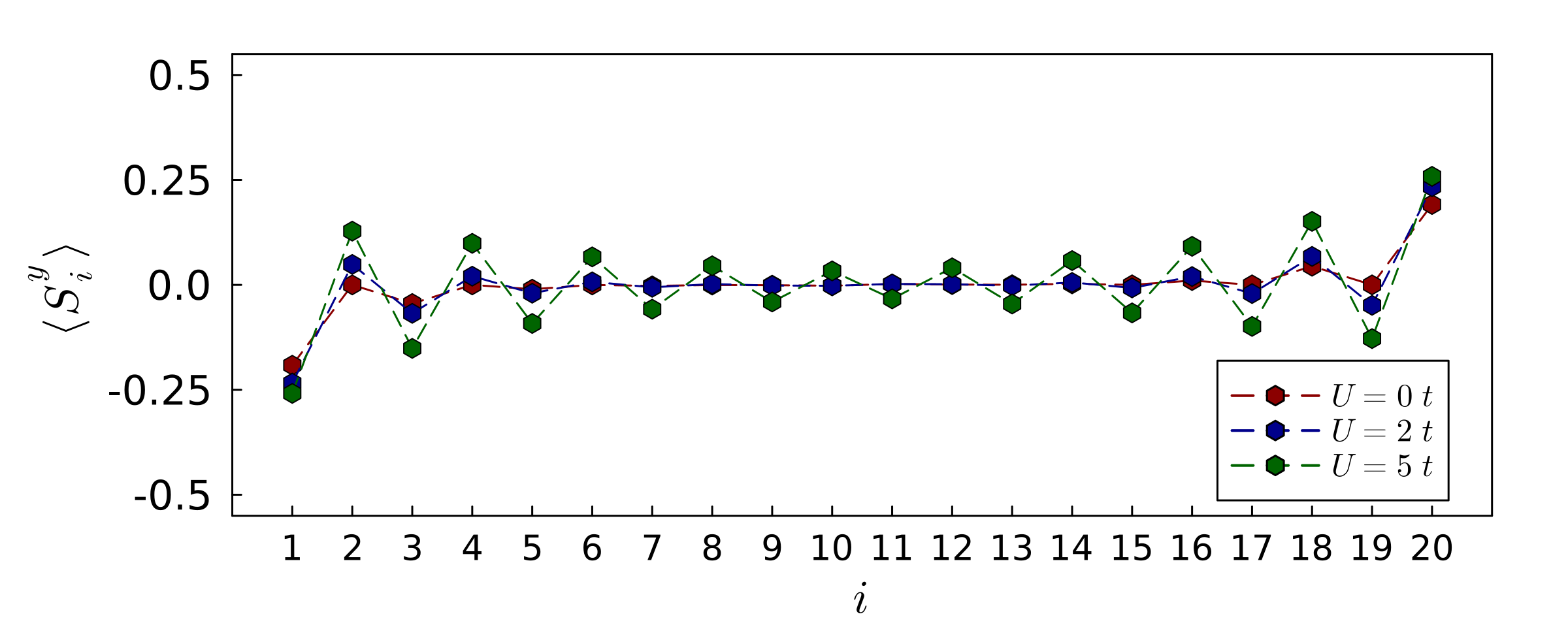}
    \caption{(Color online) Expectation value of the spin projection as 
    a function of site for a chain of 20 sites, $J_y=10 E(0)$  and different values of $U$. Other parameters as in Fig. \ref{per0}.}
    \label{syi}
\end{figure}

In Fig. \ref{e0u} we show the dependence of $E(0)$ with $U$.  The curve follows an exponential behavior with very small values for $U\to 0$ and increases abruptly for $U \sim 4$. Again, this effect can be qualitatively explained in terms of a longer localization length $\xi \sim \hbar v_F/\Delta_s $ due to the renormalization of $\Delta_s$ to lower values by the effect of the interaction. Presumably, at the value of $U\simeq 4$ the regime $L \sim \xi $ is reached, and the mixing of MZMs at different ends becomes important. We illustrate this effect in Fig. \ref{syi}, where we show the expectation 
value of the spin at each site for a chain of $L= 20$ sites. Note that for $U=0$ and $U=2$ the expectation values of $\langle S_i^y \rangle$ are localized near the ends and vanish exponentially fast near the middle of the chain. However, for  $U=5$, the magnetization is spread all over the system, indicating that the MZM localization length $\xi$ is of the order of $L$.

\begin{figure}[h]
    \centering
\includegraphics[width=0.8\columnwidth]{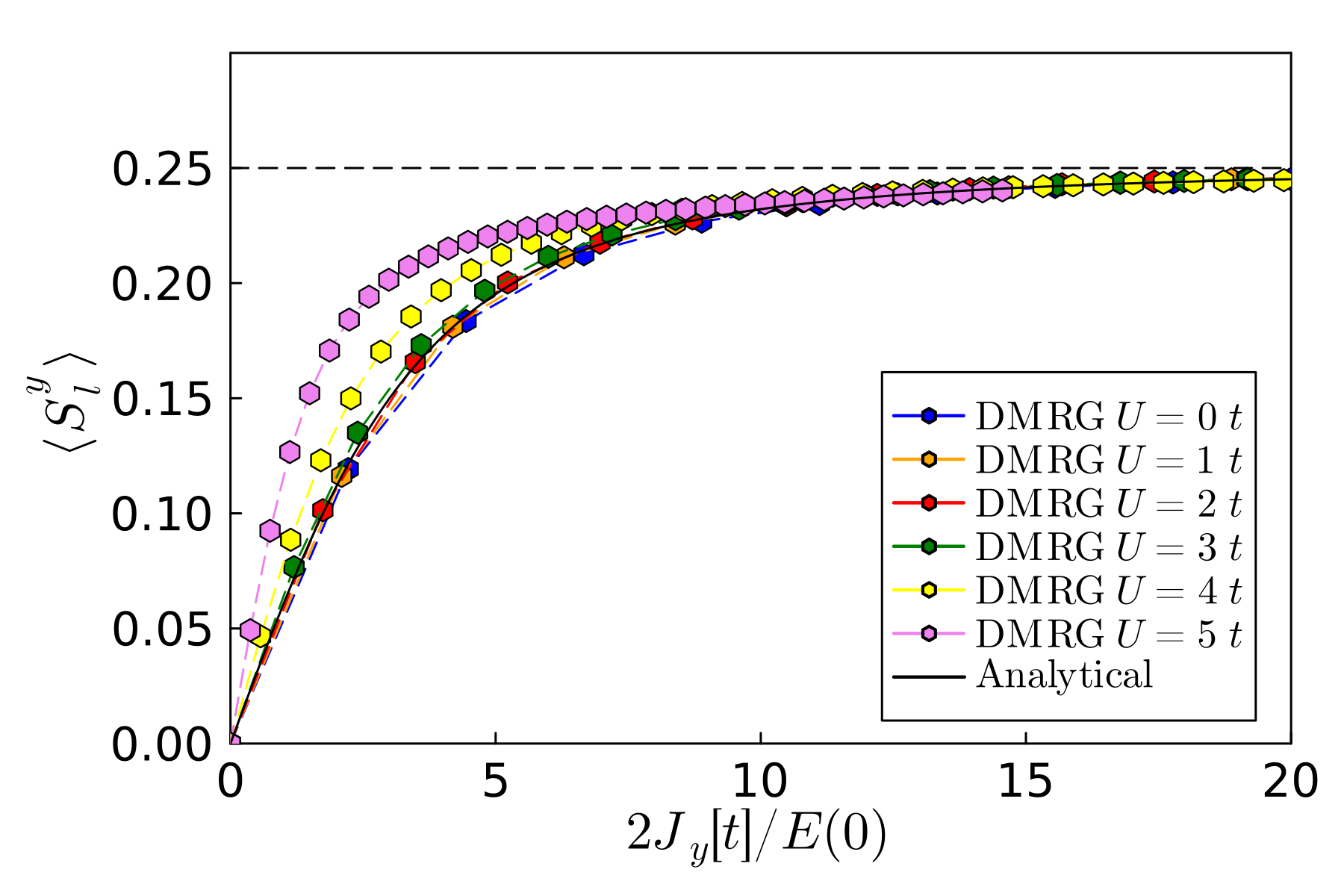}
    \caption{(Color online) Magnetization at the left side of the chain as 
    a function of magnetic field applied to the left part 
    of the chain for 20 sites and several values of $U$.
    Other parameters as in Fig. \ref{per0}.}
    \label{parau}
\end{figure}

Finally, in Fig.  \ref{parau} we show the expectation value of the spin projection at the left end as a function of the magnetic field applied parallel to the SOC, for several values of $U$. Interestingly, note that despite the fact that  Eq. (\ref{sy}) was analytically obtained for a non-interacting model, it remains valid even in the strongly-interacting regime (i.e., up to $U\lesssim 3$) and shows universal behavior. Up to  
$U\sim 5$ the expression is only  qualitatively valid and it 
eventually breaks down for $U > 7$. This deviation and breakdown at extremely large values of $U$ occurs because the MZMs (and therefore, the magnetization) are no longer localized at the ends, and the analytic approach of Ref. \onlinecite{ali18},
which assumes localized zero-energy modes at the ends, 
is no longer  valid. 

\section{Summary and discussion}

\label{sum} 

 We have studied the strongly-interacting version of the one-dimensional Zhang-Kane-Mele model for time-reversal invariant topological superconductors. The model contains extended $s$-wave superconductivity and  Rashba spin-orbit coupling (SOC) as key ingredients, and we incorporate an on-site Coulomb repulsion $U$ and an external Zeeman field in order to study the stability of the TRITOPS phase and the emerging MZMs against the combined effects of the repulsive electron-electron interaction and the external field which breaks the time-reversal symmetry. The model is relevant to understand the effect of repulsive interactions in different one-dimensional systems predicted to host TRITOPS phases, such Shiba chains on top of high-$T_c$ superconductors \cite{Crawford20_MZM_in_high_Tc_hybrids}. 

Using the DMRG technique, we have determined the 
different topological phases of the model as a function
of chemical potential and magnetic field perpendicular to
the SOC. Remarkably, the four-fold degeneracy characteristic of the 
TRITOPS phase remains stable up to quite large values of the external Zeeman field (i.e., comparable to the bandwidth). For larger values of the magnetic field only the 
two-fold degenerate topological D phase and the trivial phase
persist. 

Concerning the effect of $U$, an important conclusion of this work is that despite its effect on the topological phase diagram (i.e., redefining the topological phase boundaries), the presence of local repulsive interactions has no other qualitative effects. In fact, our results support a phenomenological picture where electron-electron interaction can be introduced in the renormalized parameters of  an effectively non-interacting model. This has been confirmed by the fact that all physical properties seem to depend on the renormalized single-particle excitation gap $\Delta_s$ (see Fig. \ref{gaps}). In few words, the interaction $U$ weakens the four-fold degenerate phase against a perpendicular magnetic field, but it favors and stabilizes this phase with respect to a varying chemical potential. From a practical perspective, this last effect could be useful in potential applications in order to  enlarge the range of chemical potential for which the topological phases exist. 

We note that the persistence of the MZMs when the interaction 
is turned on, is not a general result. For example,
in the interacting Su-Schrieffer-Heeger model, although
in presence of the on-site repulsion $U$, two different
topological sectors can still be identified by 
many-body topological invariants \cite{gura11,man12,aa23},
the MZMs end states disappear even in the topological phase for finite $U$ \cite{man12}.
The presence of zero-energy edge modes dictated by the 
bulk-boundary correspondence is modified by the possible presence of zeros of the interacting Green’s function at zero energy \cite{gura11,man12,per23}. 
Interestingly, a zero of this kind is responsible for
a topological transition in a two-channel 
spin-1 Kondo model with easy-plane anisotropy \cite{zitko21}, which explains several relevant 
experiments \cite{zitko21,ble1,ble2}.

In order to characterize the different phases and the effect of the on-site interaction, we have calculated the magnetization at one end of the 
chain when a magnetic field is applied to that end. While external magnetic fields with atomic spatial resolution are not experimentally feasible, in the case of Shiba chains spin-polarized STM tips have allowed to study the spin response of atomic chains near the ends \cite{Jeon17_Distinguishing_MZMs}. In our work, we obtain a highly anisotropic  response depending on the orientation of the externally applied field with respect to the SOC axis. For a magnetic field $J$ parallel to the 
SOC, the spin projection at the end rapidly saturates (i.e., for field strengths corresponding
to the small energy scale $E(0)$ arising from mixing and energy-splitting of the MZMs at different ends) to 
the unconventional value $S=1/4$. This splitting decays
exponentially with $L$, the length of the chain. For moderate 
values of $U$, the magnetization at the end follows
a universal curve as a function of the ratio between
magnetic field and $E(0)$
[see Eq. (\ref{sy})].

For magnetic field perpendicular to the SOC, the 
magnetization at the end increases linearly with the applied
field with a slope inversely proportional to the superconducting
gap $\Delta_s$.

\section*{Acknowledgments}

AAA acknowledges financial support provided by PICT 2018-01546 and PICT 2020A-03661 of the Agencia I+D+i, Argentina. CJG and LMC acnowledge financial support 
provided by PIP 2021-3220 of CONICET. AML acknowledges financial support from Agencia I+D+i through grant PICT-2017-2081.

\bibliographystyle{apsrev}

\end{document}